# Tunneling induced transparency and controllable group velocity in triple and multiple quantum-dot molecules

**Si-Cong Tian[1,†], Cun-Zhu Tong[1,*], Ren-Gang Wan[2], Yong-Qiang Ning[1], Li-Jun Wang[1]**

[1] *State Key laboratory of Luminescence and Applications, Changchun Institute of Optics, Fine Mechanics and Physics, Chinese Academy of Sciences, Changchun 130033, China*

[2] *State Key Laboratory of Transient Optics and Photonics, Xi'an Institute of Optics and Precision Mechanics, Chinese Academy of Sciences, Xi'an, 710119, China*

*Corresponding author:* [†] *tiansicong@ciomp.ac.cn*

[*] *tongcz@ciomp.ac.cn*

## ABSTRACT

We analyze the interaction of a triple quantum dot molecules controlled by the tunneling coupling instead of coupling laser. A general analytic expression for the steady-state linear susceptibility for a probe-laser field is obtained and we show that the system can exhibit two transparency windows. The group velocity of the probe-laser pulse is also analyzed. By changing the tunneling couplings, two laser pulses with different central frequency can propagate with the same group velocity. And the group velocity can be as low as 300 *m/s* in our system. We extend our analysis to the case of multiple quantum dot molecules (the number of the quantum dots is *N*) and show that the system can exhibit at most *N-1* transparency windows.

And at most *N-1* laser pulses with different central frequencies can be slowed down.

# 1. INTRODUCTION

The phenomenon of electromagnetically induced transparency (EIT), which based on the laser induced atomic coherence, plays an important role in the interaction between light and matter. EIT has found numerous applications in light propagation control, light storage, enhanced nonlinear optics and quantum computation and communication [1-4]. EIT has been observed in atoms [5], rare-earth-ion-doped crystals [6], ruby [7], semiconductor quantum wells (QWs) [8] and quantum dots (QDs) [9]. EIT has also been shown to occur in four-level atomic systems with various configurations [10,11] and some experimental results already exist for these systems [12,13]. Based on EIT, slow light has been demonstrated in various atomic systems [14-18] because that EIT can eliminate the absorption and refraction at the resonant frequency of a transition and greatly enhance nonlinear susceptibility in the spectral region of induced transparency of the medium. Possible applications of slow light include all-optical buffers [19], nonlinear optics with low optical intensity [20], all-optical delay of images [21], and beam splitter [22].

Fast development in the fabrication and control of mesoscopic quantum systems opens an avenue to investigate the optical analogs of a wide variety of quantum effects in condensed matter systems [23]. Being easily controllable in size and in the energy levels spacing, quantum dot molecules (QDMs) are promising candidates for the above studies. In such molecules, an external electric field allows us to control the confining potential and the number of electrons or holes, as well as their mutual interaction. For example, one can obtain transparency window induced by the tunneling in double quantum dots (DQD) [24]. And in the triple quantum dots (TQDs) [25] coupled by tunneling, it is possible to create a four-level system. TQDs can form either linear or triangular molecules and both have been realized experimentally in the last few years [26-29]. On the theoretical side, TQDs have attracted interest mostly in a triangular arrangement, the fundamental coherence phenomena, such as spin entanglers [30], coherent population trapping [31], Kondo effect [32], dark state [33] and spin-polarized currents [34]. Besides, there are also experimental and theoretical works about multiple QD system [35-38].

In this paper, we investigate the steady optical response of a TQD molecules controlled by the tunneling coupling and then the resulted slow light propagation. We use the density matrix formalism and obtain a general analytical expression for the linear susceptibility of the probe-laser field. The results show that the system can become transparent to the probe-laser field at one or two different frequencies, which is depend on the tunneling coupling. In addition, a steeper dispersion occurs at the transparency windows. And two laser pulses with different central frequency can propagate with the same group velocity. And ultra slow light can also be obtained at the ultra narrow transparency window. Therefore, to obtain a general case, the multiple QD system (the number of the QDs is $N$) is also analyzed. The results show that the system can become transparent to the probe laser at $N-1$ different frequencies. In addition, at most $N-1$ laser pulses with different central frequencies can be slowed down.

The remainder of this paper is organized as follows. In Sec. 2, we describe the TQD model and the basic density-matrix equations, and their solution for linear susceptibility of the probe-laser field and the velocity of light are derived. In Sec. 3, we describe the multiple QD model and the basic density-matrix equations, and their solution for linear susceptibility of the probe-laser field and the velocity of light are derived. Sec. 4 contains a summary and outlook.

## 2. TRIPLE QUANTUM DOT SYSTEM

The schematic representation of energy levels is shown in Fig. 1. At nanoscale interdot separation, the hole states are localized in the QD and the electron states are rather delocalized. In absence of optical excitation, there are no excitons inside all the QDs, condition represented by the state $|0\rangle$. When an laser field is applied, a direct exciton is created inside the QD 1, which corresponds to state $|1\rangle$. The external electric field modifies the band profiles alignment, allowing the electron to tunnel from QD 1 to the QD 2 and QD 3 forming the indirect excitons, which we denoted as state $|2\rangle$ and state $|3\rangle$. And the tunnel barrier in a TQDs can be controlled by placing a gate electrode between the neighboring dots. The Hamiltonian of this system in the interaction picture and in the rotating wave and dipole approximations is given by (we use units such that $\hbar = 1$)

$$H = \sum_{j=0}^{3} E_j |j\rangle\langle j| + [(\Omega_p e^{-i\omega_p t}|0\rangle\langle 1| + T_2|2\rangle\langle 1| + T_3|3\rangle\langle 1|) + \text{H.c.}], \qquad (1)$$

where $E_j = \hbar\omega_j$ is the energy of state $|j\rangle$, $\omega_p$ is the laser frequency, $\Omega_p = \boldsymbol{\mu}_{01} \cdot \mathbf{e} \cdot E$ is the Rabi frequency of the transition $|0\rangle \to |1\rangle$, with $\boldsymbol{\mu}_{01}$ being the associated dipole transition-matrix element, $\mathbf{e}$ the polarization vector and $E$ the

electric-field amplitude of the laser pulse. And $T_2$ and $T_3$ are the tunneling coupling.

We will analyze the system using a density-matrix approach. From the Liouville equation we obtain the following equations for the density-matrix elements:

$$\dot{\rho}_{01} = -i[\Omega_p(\rho_{11}-\rho_{00})+T_2\rho_{02}+T_3\rho_{03}]+(i\delta_1-\gamma_{01})\rho_{01}, \tag{2a}$$

$$\dot{\rho}_{02} = -i(T_2\rho_{01}-\Omega_p\rho_{12})+[\frac{i}{2}(\delta_1+\delta_2)-\gamma_{02}]\rho_{02}, \tag{2b}$$

$$\dot{\rho}_{03} = -i(T_3\rho_{01}-\Omega_p\rho_{13})+[\frac{i}{2}(\delta_1+\delta_3)-\gamma_{03}]\rho_{03}, \tag{2c}$$

$$\dot{\rho}_{11} = -i[\Omega_p(\rho_{10}-\rho_{01})+T_2(\rho_{12}-\rho_{21})+T_3(\rho_{13}-\rho_{31})]-\Gamma_{10}\rho_{11}, \tag{2d}$$

$$\dot{\rho}_{12} = -i[T_2(\rho_{11}-\rho_{22})-\Omega_p\rho_{02}-T_3\rho_{32}]+[\frac{i}{2}(\delta_2-\delta_1)-\gamma_{12}]\rho_{12}, \tag{2e}$$

$$\dot{\rho}_{13} = -i[T_3(\rho_{11}-\rho_{33})-\Omega_p\rho_{03}-T_2\rho_{23}]+[\frac{i}{2}(\delta_3-\delta_1)-\gamma_{13}]\rho_{13}, \tag{2f}$$

$$\dot{\rho}_{22} = -iT_2(\rho_{21}-\rho_{12})-\Gamma_{20}\rho_{22}, \tag{2g}$$

$$\dot{\rho}_{23} = -i(T_3\rho_{21}-T_2\rho_{13})+[\frac{i}{2}(\delta_3-\delta_2)-\gamma_{23}]\rho_{23}, \tag{2h}$$

$$\dot{\rho}_{33} = -iT_3(\rho_{31}-\rho_{13})-\Gamma_{30}\rho_{33}, \tag{2i}$$

$$\dot{\rho}_{ij} = -\dot{\rho}_{ji}^*, \tag{2j}$$

$$\rho_{00}+\rho_{11}+\rho_{22}+\rho_{33}=1. \tag{2k}$$

Here the detunings are defined as $\delta_1=\omega_{01}-\omega$, $\delta_2=\delta_1+2\omega_{21}$ and $\delta_3=\delta_1+2\omega_{31}$, with $\omega_{mn}$ the transition frequency between $|m\rangle$ and $|n\rangle$ states.

And $\Gamma_{10}$, $\Gamma_{20}$ and $\Gamma_{30}$ are the radiative decay rate of populations from $|1\rangle \rightarrow |0\rangle$, $|2\rangle \rightarrow |0\rangle$ and $|3\rangle \rightarrow |0\rangle$, and $\gamma_1$, $\gamma_2$ and $\gamma_3$ are the pure dephasing rate. Thus the coherence decay rate $\gamma_{mn}$ between level $|m\rangle$ to level $|n\rangle$ can be obtained

$$\gamma_{0n}=\gamma_{n0}=\frac{1}{2}(\Gamma_{n0}+\gamma_n), \text{ (n=1,2,3)} \tag{3a}$$

$$\gamma_{mn}=\gamma_{nm}=\frac{1}{2}(\Gamma_{m0}+\Gamma_{n0}+\gamma_m+\gamma_n). \text{ (m} \neq \text{n; m, n=1,2,3)} \tag{3b}$$

We assume that the system is in its ground state $|1\rangle$ for time $t=0$, i.e., $\rho_{00}(0)=1$. In order to investigate the absorption and dispersion properties of a weak probe-laser field coupling states $|0\rangle$ and $|1\rangle$ we calculate the steady-state linear susceptibility, with absorption (dispersion) determined by the imaginary (real) part of the susceptibility. In our case the steady-state linear susceptibility can be expressed as

$$\chi(\delta_1) = -\frac{\Gamma_{opt}}{V}\frac{|\mu_{01}|^2}{\varepsilon_0 \Omega_p}\rho_{01}(t\to\infty), \qquad (4)$$

where $\Gamma_{opt}$ is the optical confinement factor [39], V is the volume of a single QD, and $\varepsilon_0$ is the dielectric constant. The coherence $\rho_{01}(t)$ is obtained by solving Eq. (2) using perturbation theory. We assume that the probe laser is weak so that $\rho_{00}(t)\approx 1$ for all times. We apply this approximation to Eq. (2), take the steady-state limit and solve for $\rho_{01}(t)$ to first order in $\Omega$. The linear susceptibility then reads

$$\chi(\delta_1) = -\frac{\Gamma_{opt}}{V}\frac{|\mu_{01}|^2}{\varepsilon_0}\frac{1}{\delta_1+i\gamma_{01}-T_2^2/(\delta_1-\omega_{21}+i\gamma_{02})-T_3^2/(\delta_1-\omega_{31}+i\gamma_{03})}, \qquad (5)$$

The susceptibility goes to zero when $\delta_1-\omega_{21}=0$ or $\delta_1-\omega_{31}=0$. If $\omega_{21}\neq\omega_{31}$, then the system will become transparent at two different frequencies of the probe field. Therefore, if $\omega_{21}=\omega_{31}$, the susceptibility reduces to a form similar to that of a double QD system.

In this model, we can determine the light group velocity according to $v_g = c/[n+\omega(dn/d\omega)]$ [40], and then

$$v_g = \frac{c}{1+\frac{1}{2}\text{Re}(\chi)+\frac{\omega_p}{2}\frac{d\,\text{Re}(\chi)}{d\delta_1}}, \qquad (6)$$

where c is the light speed in vacuum.

For our investigation, the value of the radiative decay rate of populations and the pure dephasing rate are taken from Ref. [41]. The other parameters such as the optical confinement factor, momentum matrix element, and QD volume can be found in Ref. 39. And the tunneling couplings, which can be controlled by the barrier characteristics and the external electric field, are selected from Ref. [42].

With this consideration, we investigate the behavior of the absorption profile and refractive index of the medium in presence of the two tunneling couplings by solving numerically the coupled Eq. (5) in the steady regime. And the tunneling couplings can be controlled by the barrier characteristics and the external electric field.

First, we consider the case of $\omega_{21} = \omega_{31}$. The real (red dotted line) and imaginary (blue solid line) parts of the optical susceptibility obtained numerically as a function of $\delta_1$ is shown in Fig. 2, which are similar to the results obtained in a double QD system[24]. When the electric field is set as $\omega_{31} = 0$, we can see that one transparency window occurs at the position of $\delta_1 = 0$ with a symmetrical profile [blue solid line in Fig. 2(a)]. While the dispersion curve [red dotted line in Fig. 2(a)] shows a sharp variation around $\delta_1 = 0$, which represents a large positive derivative in the refractive index. In Fig. 2(b), we show the case of $\omega_{31} \neq 0$. The position of the transparency window occurs for $\delta_1 = -\omega_{31}$ with a unsymmetrical profile (blue solid line), and the dispersion curve also shows a sharp variation around $\delta_1 = -\omega_{31}$ (red dotted line). We note finally from Eq. (6) that the slow light is attained with $[d\,\text{Re}(\chi)/d\omega] > 0$. So we can attain the slow light signals when their frequencies fall into the transparency windows accompanied by the dispersion.

Next, we consider the case of $\omega_{21} \neq \omega_{31}$. The real (red dotted line) and imaginary (blue solid line) parts of the optical susceptibility obtained numerically as a function of $\delta_1$ is shown in Fig. 3. When $\omega_{21} = -\omega_{31}$, we have two transparency windows at the position of $\delta_1 = -\omega_{21}$ and $\delta_1 = -\omega_{31}$ with a symmetrical configuration [blue solid line in Fig. 3(a)]. That is the transparency window obtained in a double QD system splits into two transparency windows. Simultaneously, the dispersion curves [red dotted line in Fig. 3(a)] present a same high slope at the center of the two transparency windows, which can make the group velocity be considerably slowed down. And this symmetrical spectrum can lead to the same group velocity at two different frequencies, thus two weak pulses with different central frequencies can propagate with the same group velocity. By changing the electric field, the absorption spectrum can be made different at the two transparency windows, as displayed in Fig. 3(b), where we present the absorption for an unsymmetrical configuration with $\omega_{21} \neq -\omega_{31}$. The two transparency windows located at the position of $\delta_1 = -\omega_{21}$ or $\delta_1 = -\omega_{31}$ can be made very narrow by choosing the suitable parameters. As can be seen, one dispersion curve presents a much higher slope than the other one at their center of the transparency windows. Thus we can control the two weak pulses with different central frequencies propagating with different group velocities. And by choosing the suitable parameters, an ultra-slow light can be obtained.

The higher the slope of the dispersion is, the slower the propagating of light is. To obtain much higher slope of the dispersion curves, we need ultra-narrow transparency windows. By tuning the electric field and tunneling coupling, the transparency windows can be made very narrow. Using the material parameters in Ref. 39, we calculate the group velocity on Eq. (6), and the group velocity can be as low as 300 m/s in our system.

In order to obtain more general situation, we show in Fig. 4 the real and imaginary parts of the optical susceptibility obtained numerically as a function of the detuning parameters $\delta_1$ and $\omega_{31}$. When $\omega_{21} = \omega_{31}$, we can see from Fig. 4(a) that there is always one transparency window occurring and the position of the transparency window follows the condition $\omega_{31} + \delta_1 = 0$. The behavior of the real part of the optical susceptibility, Fig. 4(b), shows the same features along the line defined by $\omega_{31} + \delta_1 = 0$. While in the case of $\omega_{21} \neq \omega_{31}$, we can see from Fig. 4(c) that the absorption curve displays two transparency windows at the position of $\delta_1 = -\omega_{21}$ and $\delta_1 = -\omega_{31}$. And also, the dispersion curves present a very high slope at the center of the two transparency windows, as shown in Fig. 4(d). These results show that the transparency window, as well as the positive derivative in the refractive index can be controlled by changing the electric field ($\omega_{21}$ or $\omega_{31}$) and adjusting the laser detuning, such as the condition $\delta_1 + \omega_{21} = 0$ or $\delta_1 + \omega_{31} = 0$ is fulfilled.

In addition, as can be seen from Fig. 5, the absorption peak between $\delta_1 = \omega_{21}$ and $\delta_1 = \omega_{31}$ can become very narrow if the tunneling coulping $T_2$ and $T_3$ are increased [see Fig. 5(a)] or the value of $|\omega_{21}| - |\omega_{31}|$ are decreased [see Fig. 5(b)], while the outboard one near $\delta_1 = \omega_{21}$ or $\delta_1 = \omega_{31}$ can become very narrow if we choose a large value of $\omega_{21}$ or $\omega_{31}$ [see Fig. 5(c)]. Such dynamically controllable narrow gain lines may have potential applications in the accurate spectroscopic measurement.

## 3. MULTIPLE QUANTUM DOT SYSTEM

To obtain a general case, the multiple QD system (the QD number is $N$) is also analyzed. The schematic diagram of energy levels is shown in Fig. 6. In absence of optical excitation, there are no excitons inside all the QDs, condition represented by the state $|0\rangle$. When an laser field is applied, a direct exciton is created inside the QD 1, which corresponds to state $|1\rangle$. The external electric field modifies the band profiles alignment, allowing the electron to tunnel from QD 1 to the others forming the indirect excitons, which we denoted as state $|N\rangle$. The Hamiltonian of this system in the interaction picture and in the rotating wave and dipole approximations is given by (we use units such that $\hbar = 1$)

$$H = \sum_{j=0}^{3} E_j |j\rangle\langle j| + [(\Omega_p e^{-i\omega_p t}|0\rangle\langle 1| + \sum_{n=2}^{N} T_n |n\rangle\langle 1|) + \text{H.c.}], \tag{7}$$

where $T_n$ is the tunneling coupling.

We will analyze the system using a density-matrix approach. From the Liouville equation we obtain the following equations for the density-matrix elements:

$$\dot{\rho}_{01} = -i[\Omega_p(\rho_{11} - \rho_{00}) + \sum_{n=2}^{N} T_n \rho_{0n}] + (i\delta_1 - \gamma_{01})\rho_{01}, \tag{8a}$$

$$\dot{\rho}_{0n} = -i(T_n \rho_{01} - \Omega_p \rho_{1n}) + [\frac{i}{2}(\delta_1 + \delta_n) - \gamma_{0n}]\rho_{0n}, \tag{8b}$$

$$\dot{\rho}_{1n} = -i[T_n(\rho_{11} - \rho_{nn}) - \Omega_p \rho_{0n} - \sum_{m \neq n} T_m \rho_{mn}] + [\frac{i}{2}(\delta_n - \delta_1) - \gamma_{1n}]\rho_{1n}, \tag{8c}$$

$$\dot{\rho}_{mn} = -i(T_n \rho_{m1} - T_m \rho_{1n}) + [\frac{i}{2}(\delta_n - \delta_m) - \gamma_{mn}]\rho_{mn}, \tag{8d}$$

$$\dot{\rho}_{11} = -i[\Omega_p(\rho_{10} - \rho_{01}) + \sum_{n=2}^{N} T_n(\rho_{1n} - \rho_{n1})] - \Gamma_{10}\rho_{11}, \tag{8e}$$

$$\dot{\rho}_{nn} = -iT_n(\rho_{n1} - \rho_{1n}) - \Gamma_{n0}\rho_{nn}, \tag{8f}$$

$$\dot{\rho}_{mn} = -\dot{\rho}_{nm}^{*}, \tag{8g}$$

$$\sum_{n=0}^{N} \rho_{nn} = 1. \tag{8h}$$

Where the detunings are defined as $\delta_1 = \omega_{01} - \omega$, $\delta_n = \delta_1 + 2\omega_{n1}$, with $\omega_{n1}$ the transition frequency between the state $|n\rangle$ and $|1\rangle$. And the coherence decay rate $\gamma_{mn}$ between level $|m\rangle$ to level $|n\rangle$ are

$$\gamma_{0n} = \gamma_{n0} = \frac{1}{2}(\Gamma_{n0} + \gamma_n), \ (1 \le n \le N) \tag{9a}$$

$$\gamma_{mn} = \gamma_{nm} = \frac{1}{2}(\Gamma_{m0} + \Gamma_{n0} + \gamma_m + \gamma_n), \ (m \ne n; \ 1 \le m, n \le N) \tag{9b}$$

with $\Gamma_{n0}$ is the radiative decay rate of populations from $|n\rangle \to |0\rangle$ and $\gamma_n$ is the pure dephasing rate of level $|n\rangle$.

With the method used above, we calculate the steady-state linear susceptibility

$$\chi(\delta_1) = -\frac{\Gamma_{opt}}{V}\frac{|\mu_{01}|^2}{\varepsilon_0}\frac{1}{\delta_1 + i\gamma_{01} - \sum_{n=2}^{N} T_n^2/(\delta_1 - \omega_{n1} + i\gamma_{0n})}. \tag{10}$$

The susceptibility goes to zero when $\delta_1 - \omega_{n1} = 0$ with n=2 ~ N. Therefore, if all $\omega_{n1}$ are different, then this multiple QD system will become transparent at $N-1$ different frequencies of the probe field.

In the case of $L-1$ (2<L<N) of $\omega_{n1}$ are equal to $\omega$, and the remaining $N-L$ are different from $\omega$. To simplify the notation, we take $\omega_{21} = \omega_{31} = \cdots = \omega_{L1} = \omega$. Then the susceptibility becomes

$$\chi(\delta_1) = -\frac{\Gamma_{opt}}{V}\frac{|\mu_{01}|^2}{\varepsilon_0}\frac{1}{\delta_1 + i\gamma_{01} - \sum_{n=2}^{L} T_n^2/(\delta_1 - \omega + i\gamma_{0n}) - \sum_{n=L+1}^{N} T_n^2/(\delta_1 - \omega_{n1} + i\gamma_{0n})}. \tag{11}$$

From Eq. (11), we can see that there are $N-L+1$ transparency windows in the system.

Finally, if all $\omega_{n1}$ are equal to $\omega$ then the susceptibility reduces to

$$\chi(\delta_1) = -\frac{\Gamma_{opt}}{V}\frac{|\boldsymbol{\mu}_{01}|^2}{\varepsilon_0}\frac{1}{\delta_1 + i\gamma_{01} - \sum_{n=2}^{N} T_n^2/(\delta_1 - \omega + i\gamma_{0n})}. \quad (12)$$

In this case, only one transparency window can be obtained, with the magnitude squared of the tunneling coupling is replaced with the sum of the magnitude squared of all the tunneling coupling.

The group velocity of the probe pulse can also be expressed by Eq. (6) in terms of multiple QD system. When none of $\omega_{n1}$ are the same, the group velocity at the $nth$ transparency window approximates

$$v_g = \frac{c|T_n|^2 V\varepsilon_0}{2\pi\omega_p \Gamma_{opt}|\boldsymbol{\mu}_{01}|^2}, \text{ (n=2 ~ N)}, \quad (13)$$

therefore the group velocity of the probe-laser pulse may be significantly reduced. And the group velocity can be controlled via the intensity of the tunneling couplings and the probe-laser field can propagate with $N-1$ different group velocities in the medium.

When $L-1$ (2<L<N) of $\omega_{n1}$ are equal to $\omega$ (to simplify the notation, we take $\omega_{21} = \omega_{31} = \cdots = \omega_{L1} = \omega$), then the group velocity of the probe pulse becomes the same as Eq. (13) at the $nth$ transparency window center with $n = L+1$, ... , $N-1$, and

$$v_g = \frac{c\sum_{n=2}^{L}|T_n|^2 V\varepsilon_0}{2\pi\omega_p \Gamma_{opt}|\boldsymbol{\mu}_{01}|^2}, \quad (14)$$

around the other transparency window. Finally if all $\omega_{n1}$ are equal to $\omega$, then the group velocity approximates

$$v_g = \frac{c\sum_{n=2}^{N}|T_n|^2 V\varepsilon_0}{2\pi\omega_p \Gamma_{opt}|\mathbf{\mu}_{01}|^2},\tag{15}$$

around the single transparency window.

We will now give one example of the steady optical response that could occur in multiple QD system. We show in Fig. 7 the absorption and dispersion spectra of the five QD molecules. The absorption and dispersion are either symmetrical or unsymmetrical and their shapes depend on the system parameters. In the case of $\omega_{21} \neq \omega_{31} \neq \omega_{41} \neq \omega_{51}$, four transparency windows occur in Fig. 7(a) and 7(b). When all the tunneling couplings are equal [Fig. 7(a)], the spectrum is symmetrical, supporting the propagation of four weak laser pulses with the same group velocity at different central frequencies. Otherwise, the spectrum is unsymmetrical and four weak laser pulses can propagate with different group velocity, which is the case of Fig. 7(b). In Fig. 7(c) and 7(d) we show the results with two equal detunings. In this case only two transparency windows appear in the spectrum and the group velocity can obtain either different or same value depending on the tunneling coupling. Thus the weak laser pulses can propagate with the same or the different group velocity.

## 4. CONCLUSION AND OUTLOOK

In summary, we have studied the triple and multiple QD molecules under the tunneling couplings. A general analytic expression for the steady state linear susceptibility for a probe-laser field has been obtained. We have shown that the system (the number of QDs is $N$) can exhibit at most $N-1$ transparency windows. The absorption and dispersion are either symmetrical or unsymmetrical which depend critically on the tunneling couplings. Thus the group velocities can be controlled by varying the tunneling couplings. In the case of symmetrical spectrum, the weak laser pulses with different central frequencies can propagate with the same group velocity. Otherwise, the weak laser pulses can propagate with different group velocity and ultra slow light can be obtained.

These results allow us to optimize the slow light phenomena in artificial multiple QD molecules by using the electric field and tunneling coupling instead of coupling laser. And the parameter tunneling depends on the multiple QD molecules barrier, which is defined in the epitaxial growth process. Such a controllable group velocity may open the way to new interesting experiments in quantum optics and quantum information, and might be a solution not only for buffering but also various types of time-domain processing, such as retiming, multiplexing and performing convolution integrals.

Moreover, optical cavity is a useful tool for investigation of light-matter interaction. Superluminal and subluminal propagation has been experimentally demonstrated in a high-$Q$ optical microcavity containing a few cold atoms in its cavity mode [43]. In QDMs, when tunneling induced transparency is incorporated with a resonant cavity, the group velocity of the light pulse can be much slower due to the round-trip behaviour of signal light in the cavity. And thanks to the impressive recent progress, photonic crystals can probably be used to build microcavities at optical frequency [44]. Up to now, a variety of nanocavities and waveguides have been designed and fabricated for future nano-optics devices [45]. Hence in these quantum-dot-cavity systems, the group velocity of the laser pulse can be greatly slowed down.

## ACKNOWLEDGMENTS

This work is supported by the financial support from the National Basic Research

Program of China (Grant No. 2013CB933300), the National Natural Science Foundation of China (Grant No. 11304308, 61076064 and 61176046), and the Hundred Talents Program of Chinese Academy of Sciences.

# FIGURE CAPTIONS

Fig.1    Schematic of the setup of the TQDs. (a) An optical pulse transmits the QD 1. V is a bias voltage. (b) Schematic band structure and level configuration.

Fig.2    The absorption (blue solid curves) and dispersion spectra (red dotted curves) as a function of detuning $\delta_1$ for a TQDs with $T_2 = 10\mu eV$, $T_3 = 10\mu eV$, $\Gamma_{10} = 6.6\mu eV$, $\Gamma_{20} = \Gamma_{30} = 10^{-4}\Gamma_{10}$, $\gamma_{10} = 20\mu eV$, $\gamma_{20} = \gamma_{30} = 10^{-3}\gamma_{10}$, and (a) $\omega_{21} = \omega_{31} = 0$, (b) $\omega_{21} = \omega_{31} = 20\mu eV$.

Fig.3    The absorption (blue solid curves) and dispersion spectra (red dotted curves) as a function of detuning $\delta_1$ for a TQDs with (a) $\omega_{21} = 10\mu eV$, $\omega_{31} = -10\mu eV$, $T_2 = T_3 = 10\mu eV$, (b) $\omega_{21} = 10\mu eV$, $\omega_{31} = -2.5\mu eV$, $T_2 = 2.5\mu eV$, $T_3 = 10\mu eV$. Other parameters are the same as those in Fig. 2.

Fig.4    The absorption [Fig. 4(a) and 4(b)] and dispersion spectra [Fig. 4(c) and 4(d)] as a function of detuning $\delta_1$ and $\omega_{31}$ for a TQDs with (a) and (c) $\omega_{21} = \omega_{31}$, (b) and (d) $\omega_{21} = 10\mu eV$. Other parameters are the same as those in Fig. 2.

Fig.5    The absorption (blue solid curves) and dispersion spectra (red dotted curves) as a function of detuning $\delta_1$ for a TQDs with (a) $\omega_{21} = 5\mu eV$, $\omega_{31} = -5\mu eV$, $T_2 = T_3 = 20\mu eV$, (b) $\omega_{21} = 2\mu eV$, $\omega_{31} = -2\mu eV$, $T_2 = T_3 = 10\mu eV$, (c) $\omega_{21} = 50\mu eV$, $\omega_{31} = 0$, $T_2 = 10\mu eV$, $T_3 = 10\mu eV$. Other parameters are the same as those in Fig. 2.

Fig.6    Schematic of the setup of the multiple QD system. (a) An optical pulse transmits the QD 1. V is a bias voltage. (b) Schematic band structure and level configuration.

Fig.7 The absorption (blue solid curves) and dispersion spectra (red dotted curves) as a function of detuning $\delta_1$ for a multiple QD system ($N=5$) with (a) and (b) $\omega_{21}=15\mu eV$, $\omega_{31}=5\mu eV$, $\omega_{41}=-5\mu eV$, $\omega_{51}=-15\mu eV$, (c) and (d) $\omega_{21}=\omega_{31}=10\mu eV$, $\omega_{41}=\omega_{51}=-10\mu eV$. (a) and (c) $T_2=T_3=T_4=T_5=10\mu eV$, (b) and (d) $T_2=T_3=5\mu eV$, $T_4=T_5=10\mu eV$. Other parameters are $\Gamma_{10}=6.6\mu eV$, $\Gamma_{n0}=10^{-4}\Gamma_{10}$, $\gamma_{10}=20\mu eV$, $\gamma_{n0}=10^{-3}\gamma_{10}$.

# FIGURE

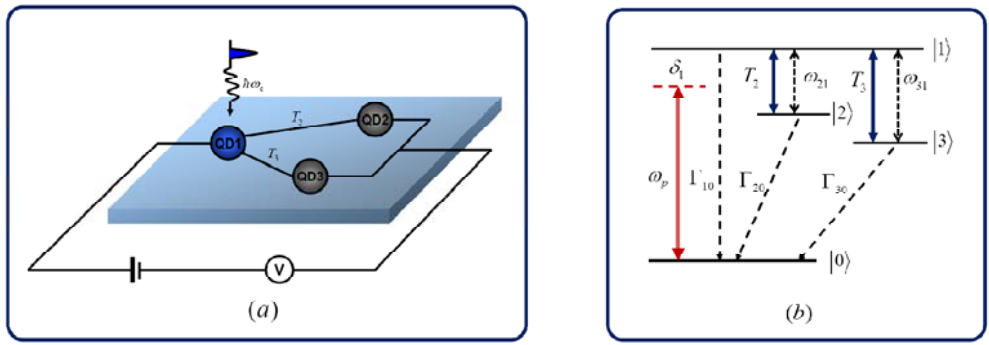

Fig. 1

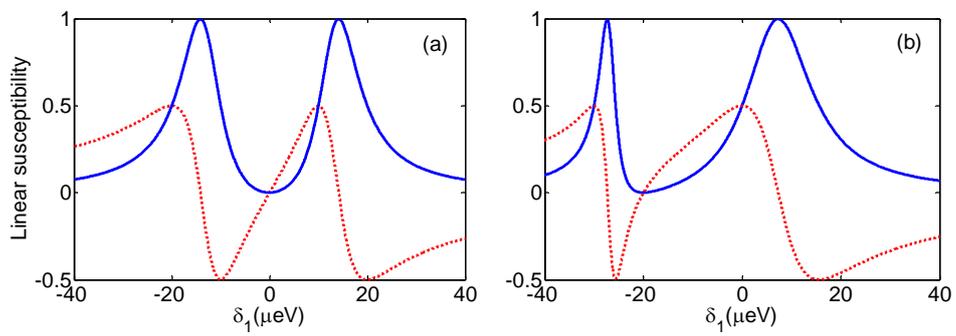

Fig. 2

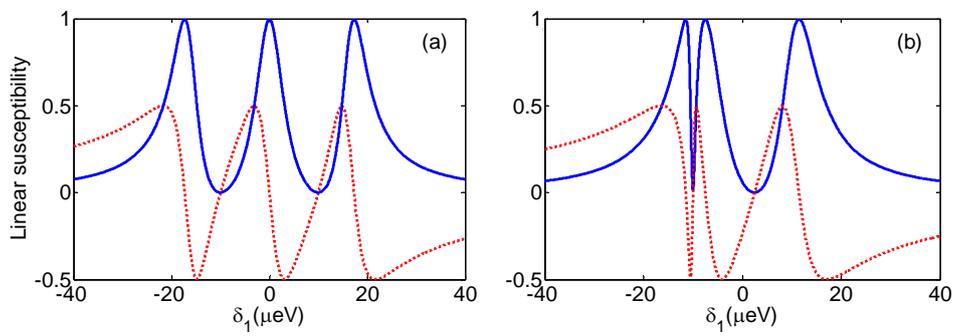

Fig. 3

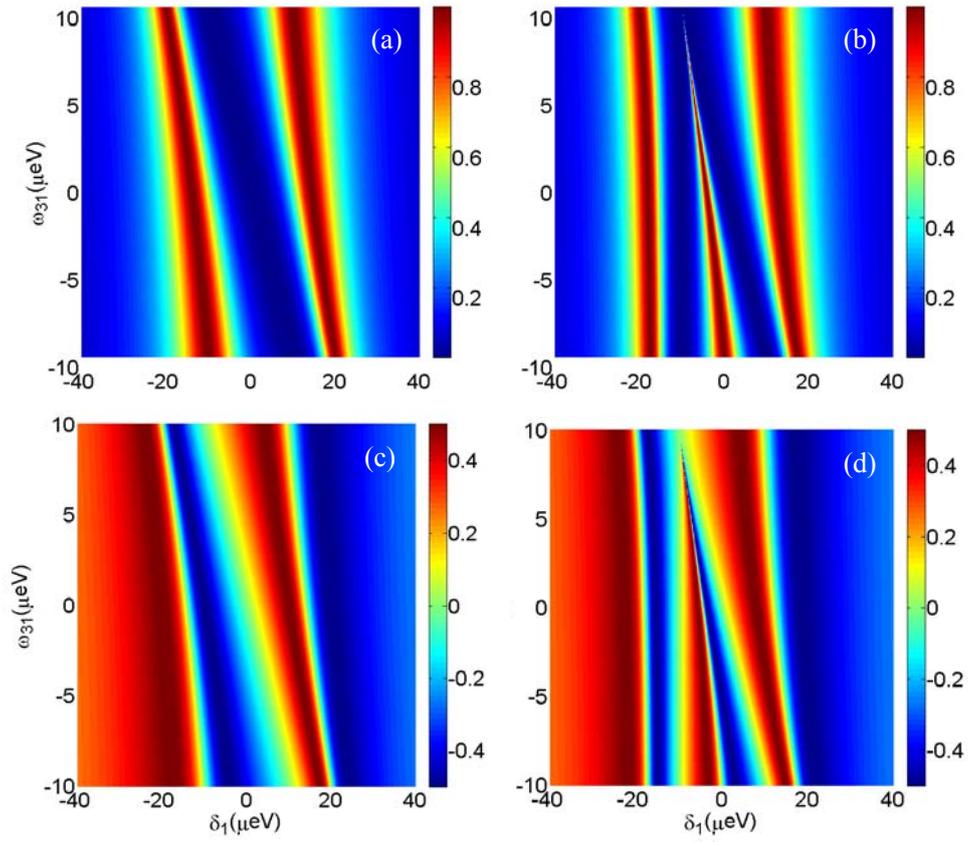

Fig. 4

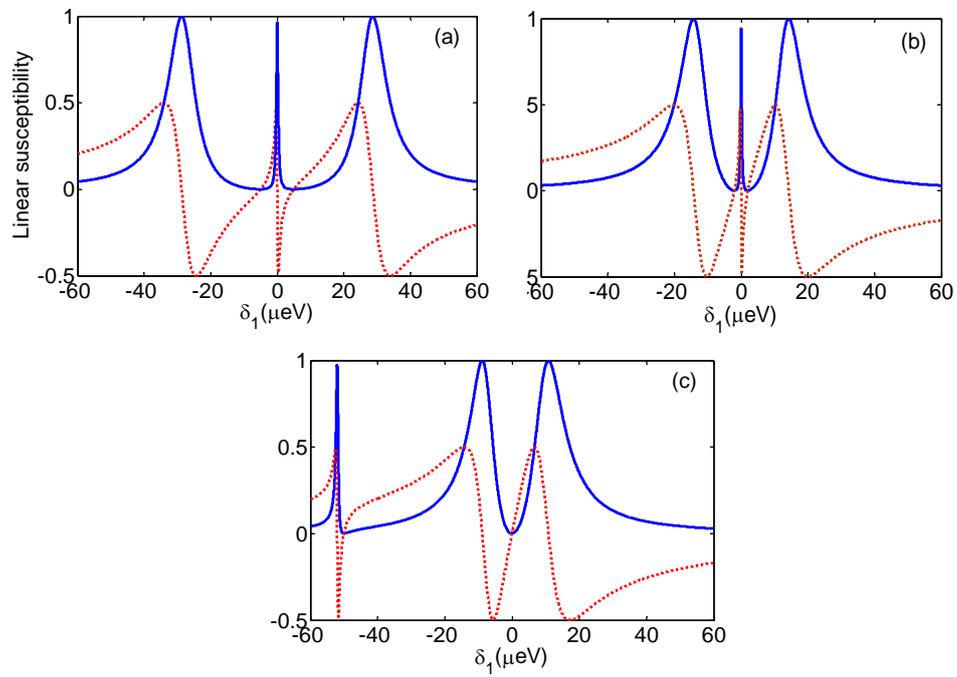

Fig. 5

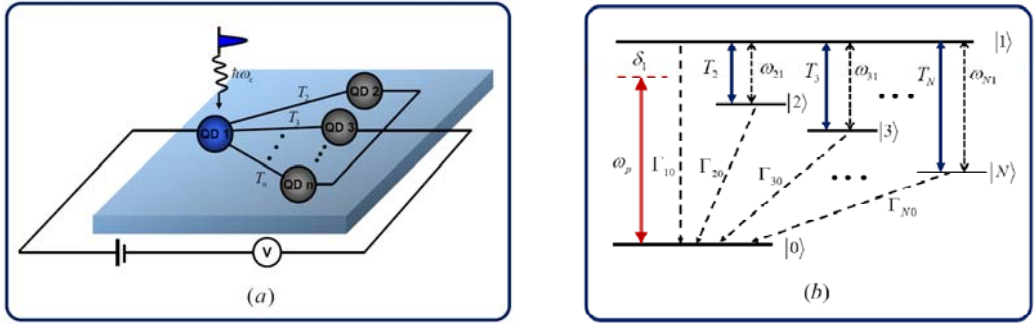

Fig. 6

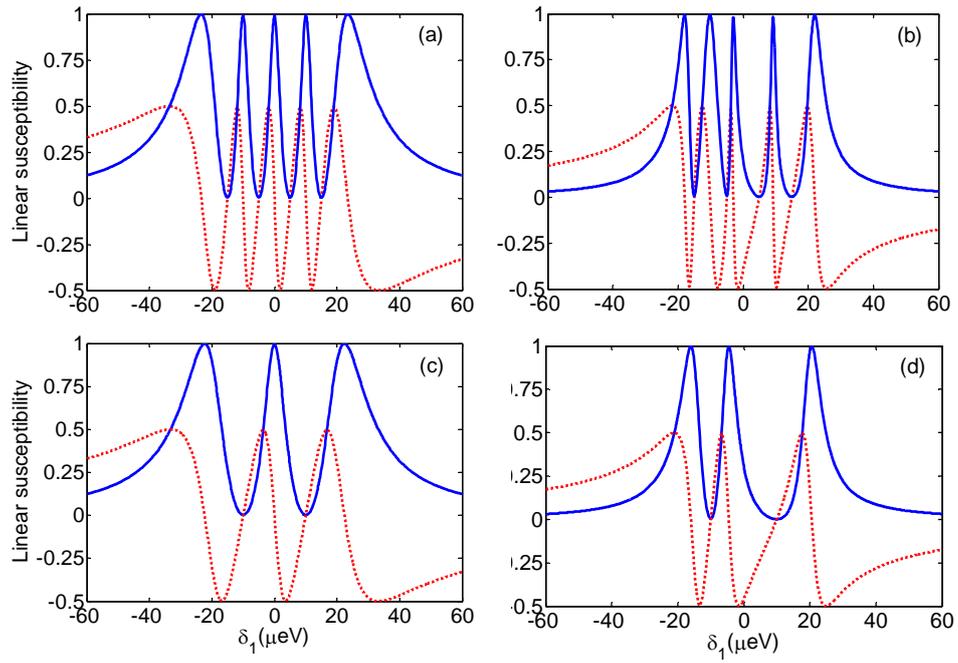

Fig. 7